# Analyzing Standardization Needs for CHIL-based Testing of Power Systems and Components

Georg Lauss, Filip Pröstl Andrén, Fabian Leimgruber, Thomas I. Strasser
Center for Energy – Electric Energy Systems
AIT Austrian Institute of Technology, Vienna, Austria
Email: {georg.lauss, filip.proestl-andren, fabian.leimgruber, thomas.strasser}@ait.ac.at

***Abstract*—Real-time simulation methods for investigations on electric networks and integration of grid connected generation units are increasingly in the focus of ongoing research areas. While laboratory testing methods are the predominant method for the verification of safety and quality related features of grid-connected generation units in the past, load flow modeling verification methods have been integrated in state-of-the-art standardization frameworks recently. The next step is comprised in real-time simulation methodologies applied for compliance testing of entire power electronic systems integrated in power distribution networks. The Controller Hardware-in-the-Loop (CHIL) approach is an appropriate methodology that combines numerical simulations with software modeling approaches and classical hardware testing in labs. Control boards represent the hardware device directly connected to the power electronic periphery, which is entirely simulated in a real-time simulation environment. Hereby, input signals from voltage and current measurements and output signals for power system control are exchanged in real-time. Thanks to this setup the testing of the true behavior of entire generation units within the electric network can be emulated precisely. With the application of CHIL a shorter time to market and a lower risk in the development phase can be achieved. However, an analysis from realized CHIL experiments shows the need for more harmonized procedures. This paper addresses this topic and provides an outlook about necessary future CHIL standardization needs.**

## I. Introduction

Large-scale integration of renewable energies such as solar leads to an increasing level of distributed power generation. Each generation plant is obliged to comply with subsequently mentioned power quality and safety standards. Methods of testing according to given standards [1]–[7] applying for grid connected inverters are increasingly in the focus of Distributed Generation (DG) research.

While various principles have been developed over the last decades related to real-time simulation techniques, nowadays Controller Hardware-in-the-Loop (CHIL) simulation [8]–[13] is becoming increasingly important for pre-certification services related to both generation unit and network investigations. The CHIL technique involves closed-loop simulation of two or more subsystems, at least one of which is a physical device and another a software-only subsystem executed on a Digital Real-Time Simulator (DRTS) platform [8]. Hardware under Test (HUT) is interfaced to the DRTS using analog or digital communication, which is sampled deterministically by the DRTS [9], [11]. The HUT is generally an embedded controller or other device with input or output using signal-level voltages. This technique offers the ability to interface with physical device controllers in the state in which they will finally be implemented. In so doing, nonidealities introduced by embedding control logic on the target platform and physical signals are included in the overall closed-loop simulation. Compared with Power Hardware-in-the-Loop (PHIL) simulation, no power amplification stage whatsoever is integrated in the CHIL methodology [8], [9], [11].

In the past, laboratories such as the SmartEST lab located at the AIT Austrian Institute of Technology provided exclusive infrastructure for high power component testing. Thanks to the CHIL methodology, hardware testing facilities can be emulated in real-time environment platforms and testing of high-power components can be facilitated with respect to flexibility in system complexity, power and voltage scalability, and automated testing. This contribution gives statements based on reproducible laboratory test measurements and comparative CHIL test results based on pre-certification testing and state-of-the-art network analysis.

Following this introduction, a thorough background of the idea of the CHIL simulation methodology and the design principles of the implementation are given in Section II. In Section III, three CHIL simulation setups related to compliance testing of generation units and entire network investigations are presented as applied testing scenarios. In Section IV, challenges in the context of advanced testing methodologies are discussed with focus on their relevance for upcoming standardization frameworks. The paper conclusions with the main findings in Section V.

## II. Advanced Testing via CHIL Method

In this Section, the basic architecture of CHIL simulation systems is characterized. Important principles of the CHIL methodology are discussed and given limitations as well as specific design rules for CHIL simulation systems are stated in the following.

The CHIL method normally refers to a HIL setup where the signals exchanged between the HUT and the DRTS are at a low power rating (e.g., at a low voltage or current level) or through a digital communication connection. Such a configuration can be used to test and validate architectures, concepts and algorithms implemented in real controller devices (i.e., HUT) [14].

Furthermore, the CHIL method is also a rapid prototyping and testing method [9]. It combines the advantages of hardware testing methods and simulation methods. The basic idea is that a controller running on its target platform is tested against a simulated version of the plant.

An example is an inverter controller running on a micro controller, which is tested against a simulation of inverter and the electrical grid. However, the concept is scalable and it is even possible to simulate a whole power distribution network, or even a transmission system. CHIL simulations are mostly executed in real-time, however also non-real-time simulations are possible. The time-frame of simulation is dependent on the controller implementation. Figure 1 shows the overview and the basic setup of the CHIL methodology using a DRTS as a simulation platform [15].

## III. Selected Testing Scenarios

In this Section, three testing scenarios with applied CHIL simulation methodology are discussed. The first scenario describes a CHIL test setup as an embedded platform for grid compliance testing of grid-connected generation units. This scenario is followed by a CHIL simulation system which targets the detailed investigation of voltage stability problems within electric power networks as well as by a CHIL-based testing of a lab automation approach.

### *A. Testing of requirements for grid-connection generators*

The driving motivation for the testing of requirements is to provide an end-to-end CHIL environment to support controller development from sketching test schematics through interactive and automated controller application up to standardized pre-certification testing and reporting.

Subsequently, the basic components for the CHIL test setup are presented providing an embedded platform with the complete test setup and fully automated testing and reporting at the end. Fig. 2 highlights the schematics of a power generation source PV1 (represented by a photo-voltaic array), followed by the input stage implemented as a boosting stage, the symmetric DC-link stage (DC+, DCn, DC-), and the T-type electronic power bridge with phases A, B, and C as the generalized generator output stage.

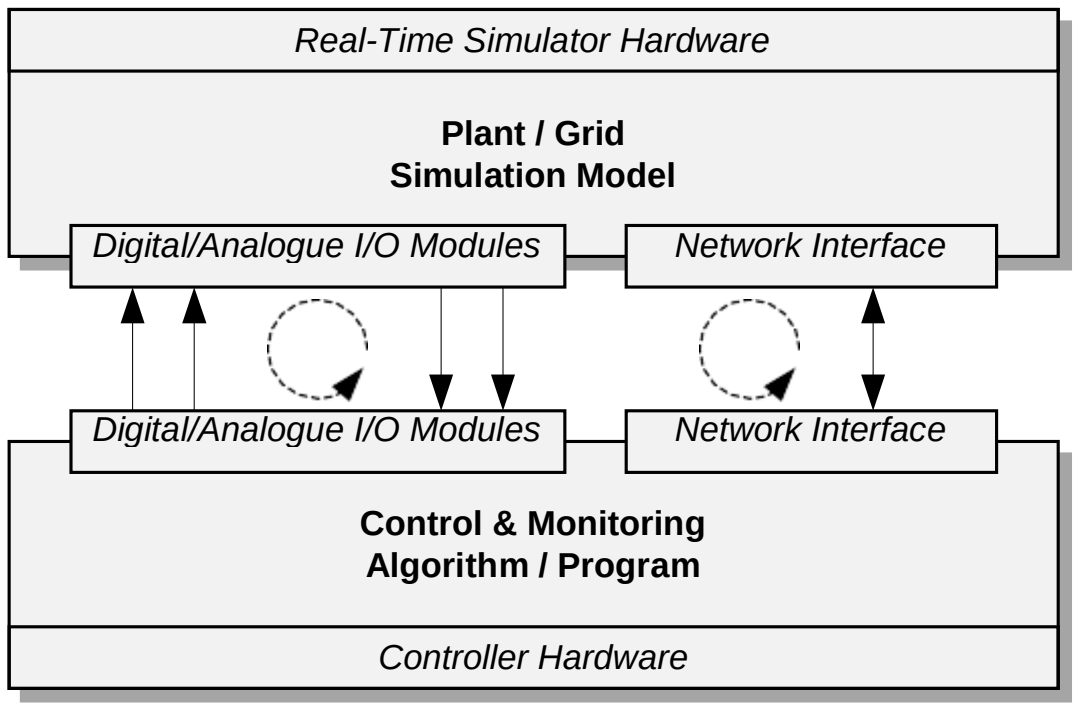

Fig. 1. Overview of the CHIL validation approach.

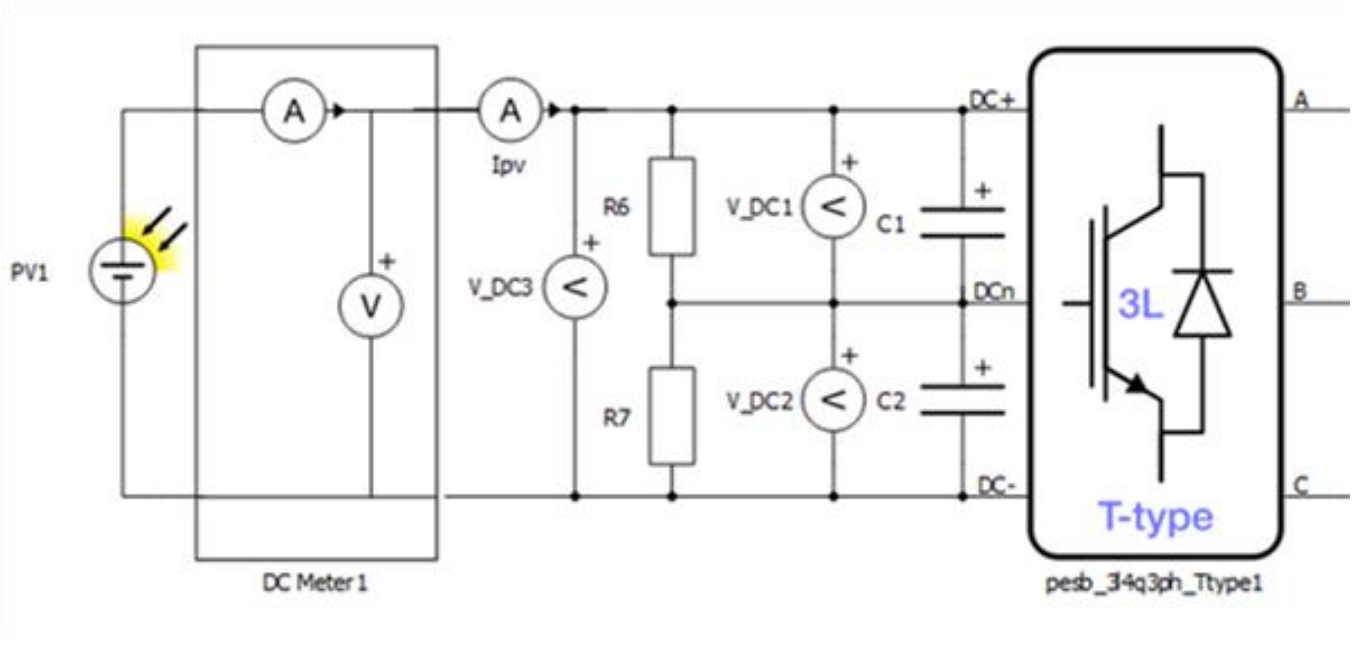

Fig. 2. Schematic view of BDEW LVRT test setup with PV source, converter and grid components selectable from a pre-built component library.

There are numerous tests included in the standards classified with required test setups, data recording, and reporting structure. In this contribution, the exemplary CHIL setup that illustrates the components, structure and dependencies will be the pre-certification testing according to FGW TR3, Rev 24 [3]. Specifically, the CHIL simulation test for item BDEW TR3, 4.7 [2], generally known as the Low-Voltage Ride-Through (LVRT) or Performance during voltage dips, is set up by

1) Specifying power circuit representing the test setup with PV simulator, HUT and grid components as built by the DRTS machine.
2) Specifying the parametrization of the mentioned components, e.g., grid simulator phase layout and frequency, PV system size and STC settings, and I/O channel setup.
3) Setting grid voltage configuration and rated power of the HUT.

This procedure accommodates for individual test setup requirements such as I/O channel setup, HUT parametrization as well as an iterative development approach. As an example, a specific CHIL simulation testing scenario is highlighted in this section to demonstrate the applicability of the HIL methodology for standardization testing frameworks such as the latest revision of the BDEW guideline.

Fig. 3 shows an entire voltage dip event and illustrates relevant 3-phase signals (axis in Fig. 3 from top to bottom: DC voltages, line-to-neutral voltage, grid currents, generator side currents) as oscilloscope signals. The signals before the fault start and after the fault end are included as well as the complete duration of the voltage dip.

In Fig. 4, the 3-phase line-to-neutral voltages U L1 (blue), U L2 (red), and U L3 (green) are shown in detail with a 200 ms time window centered to the point of time at fault start. Hereby, the transients of the line voltages can be analyzed in-depth and they can be compared to laboratory tests of hardware generation units in order to verify the CHIL methodology.

Fig. 5 highlights the 3-phase line currents I L1 (blue), I L2 (red), and I L3 (green) with the same time period settings as for Fig. 4. Analogously, the transient behavior of line currents at the point of time for the incipient fault start can be analyzed and compared to results with hardware testing scenarios.

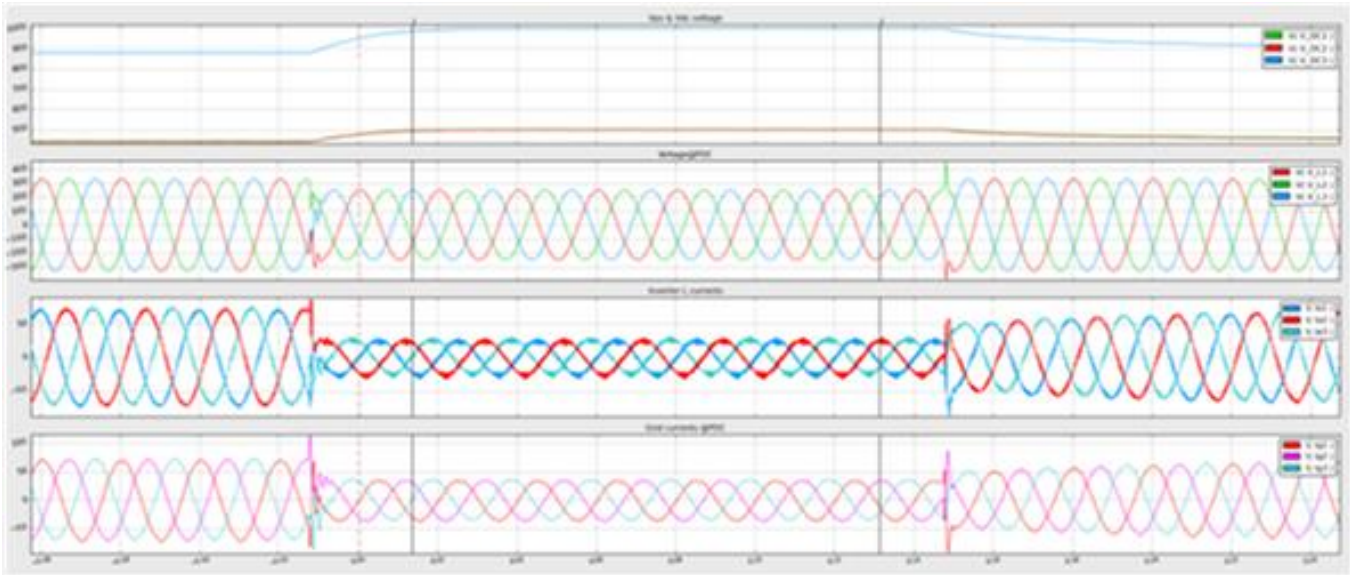

Fig. 3. Illustration of inverter and grid currents/voltages at fault start and end.

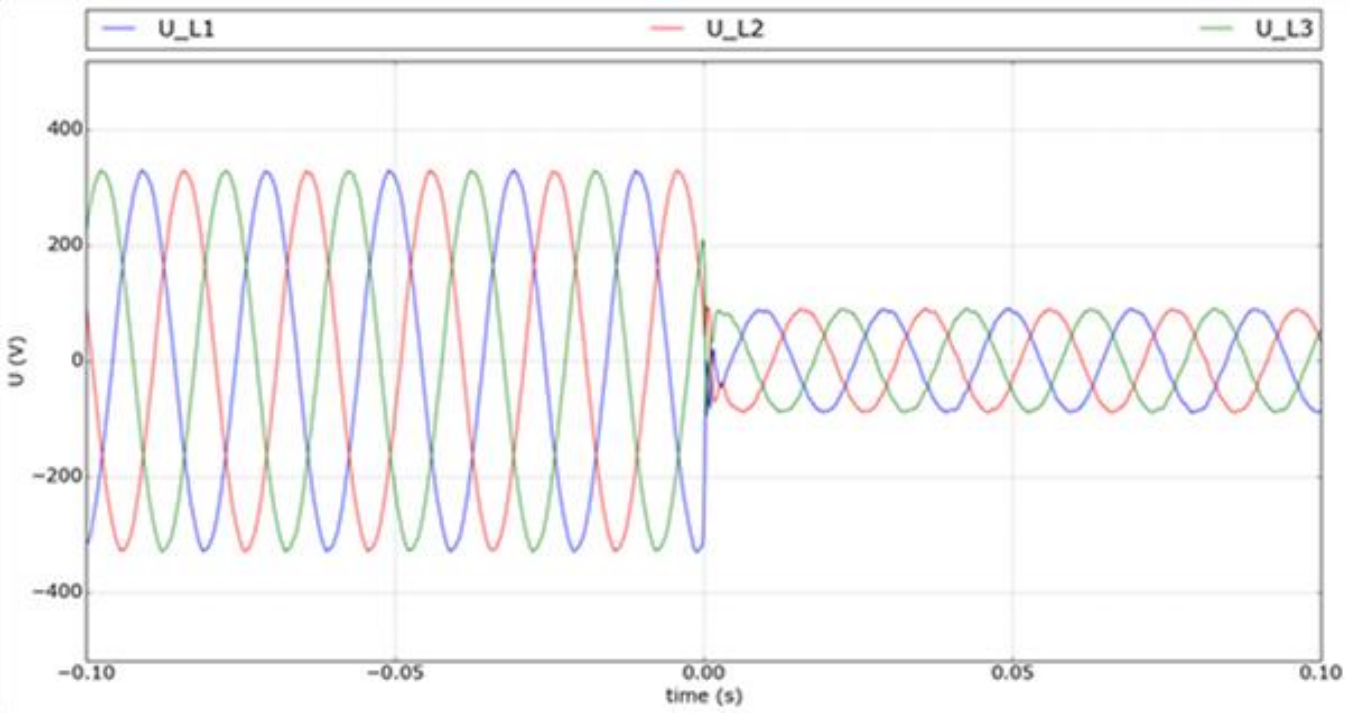

Fig. 4. Signal waveforms of 3-phase line-to-neutral voltages U_L1, U_L2, U_L3 before and after the initiated fault start at *t*=0.

Fig. 6 illustrates resulting normalized value of the positive sequence apparent current Ib 1 per norm marked with the blue solid line. The upper limit specified according to the applied standard (1.2 p.u.) is marked with by the red solid line and the lower limit (0.9 p.u.) is marked by the green solid line. A time period of 100 ms before and after the fault start is documented in order to be able to evaluate the rise time of the apparent current signal according to the given requirements.

The time step for this CHIL setup is set to 1 µs which provides high-fidelity for transient signals of the LVRT scenario. In general, the given results are compared to offline simulation and hardware experiments in order to evaluate the accuracy of the CHIL simulation. A comparison is often performed analyzing the THDV or THDI, as well as an identification of the rise times of the apparent current signal.

*B. Investigations on voltage stability of networks*

As an example test case to show how CHIL can be used for investigations of voltage stability in distribution grids the development of a Central Voltage Control Unit (CVCU) is analyzed [16]. The controller is used for a coordinated voltage control in a medium voltage network using a tap changer controlled transformer and the reactive power of controllable DGs. For CHIL testing the controller is connected with a real-time power system simulator.

The prototype of the CVCU was developed using MATLAB and tested and validated with offline simulation. Different scenarios and operation cases, especially worst cases that cannot be tested in the real network, were simulated. The obtained results were mathematically analyzed offline and the controller was improved to reach the most appropriate algorithm. These insights led to different redesigns and development iterations [16], [17].

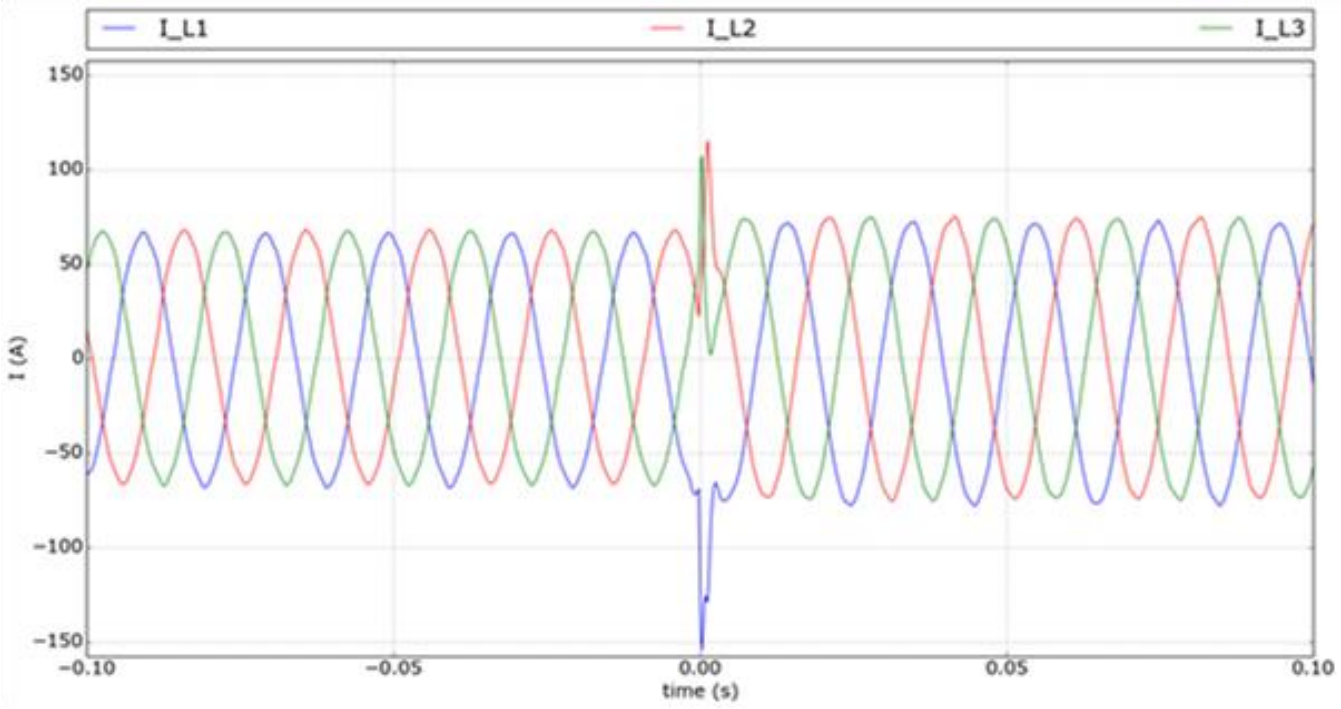

Fig. 5. Signal waveforms of 3-phase line currents fault I_L1, I_L2, I_L3 before and after the initiated fault start at *t*=0.

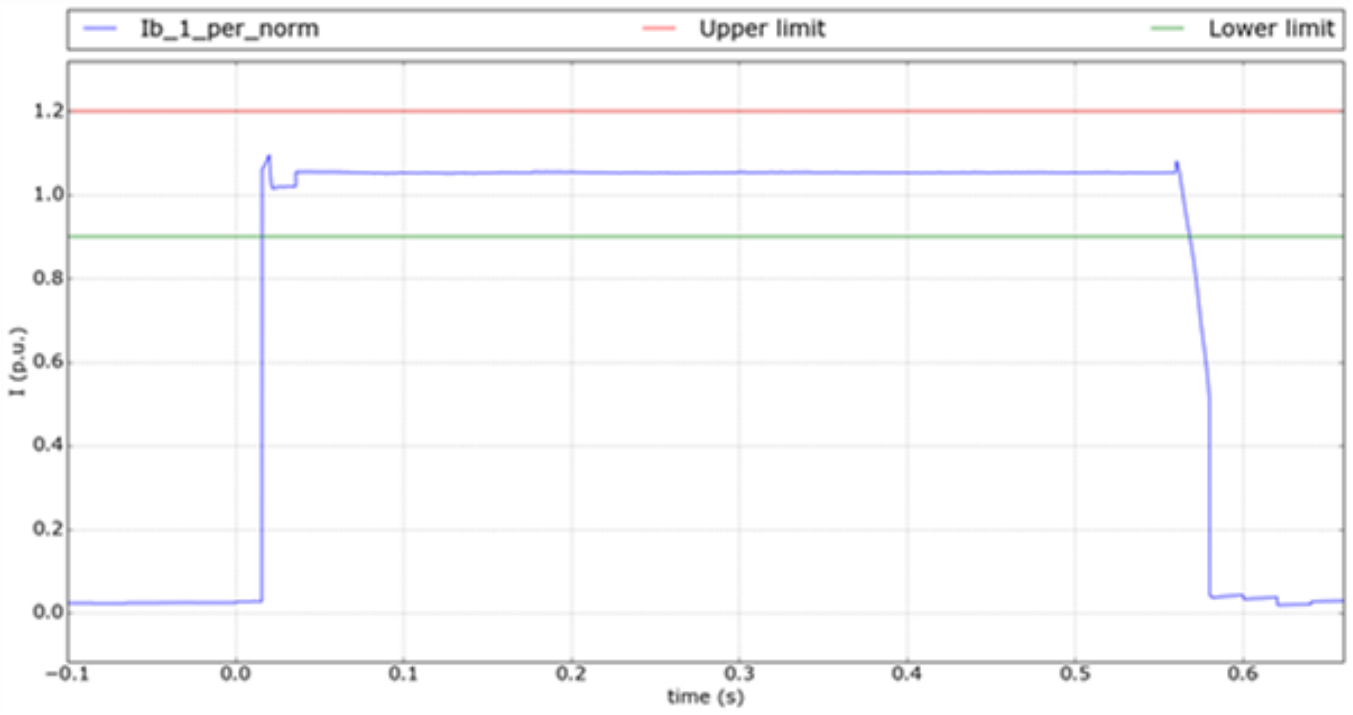

Fig. 6. Positive sequence normalized value of the apparent current Ib_1_per_norm with given upper and lower limitations.

For the realization of the CVCU a C++ implementation was chosen. The concept of the CVCU can be divided into two parts: The control of the voltage level and the control of the voltage band. The range controller keeps the spread between the highest and the lowest voltage within the allowed voltage band regardless of the absolute voltage value. This is done by optimizing the reactive power of the DGs. The level controller changes the transformer tap. This increases or decreases all voltages simultaneously to bring them back into the allowed voltage band [16].

To validate the C++ realization of the CVCU a normal offline simulation is not possible. Furthermore, to validate it directly against the real distribution system would be unsafe. The electric energy system counts as a critical infrastructure. A failure may lead to severe personal or economical damage. Since the CVCU controls the state of the distribution system it means that even a small bug in the CVCU can lead to a failure of the system and thus cause damage. This must be avoided at all cost. Using a CHIL simulation is one possibility to overcome this problem. Furthermore, a shut-down of electric supply networks include costly measures. Hence, CHIL is a cost effective method compared to tests with real systems [18].

During real operation the communication interface of the CVCU has the task of communication between the controller and SCADA system of the distribution network operator. This includes handling voltage measurements from the the grid and set points from the controller to the transformer and DGs. For the CHIL testing purposes an interface between the controller and the real-time simulator is also needed. Although these interfaces may differ, it is still a possibility to investigate how the controller operates using an external interface.

The goal of the real-time simulation was to validate the functionality of the software developed for the CVCU. The distribution network was simulated using DIgSILENT/PowerFactory which was connected to the controller using an OPC interface. Using the real-time functionality of PowerFactory the simulations were synchronizes with the system time of the computer.

During the validation process different parts of the controller could be tested, evaluated, and accordingly improved. Although the concrete control algorithms had already been validated using the prototype many new problems arise when a realization is made.

In Fig. 7 only the voltage level control of the controller was tested. The configuration was set to minimize the number of taps (i.e., the transformer tap is only changed if there is an over voltage or an under voltage). The upper and lower voltage limits used in the simulations were 1.02 and 0.94 (p.u.).

During the real-time simulation the CVCU controller was executed with a 7.5 s cycle. This simulation was first executed without communication delay and secondly with a 60 s communication delay/interruption. The magnified parts in Fig. 7 shows the duration of an over voltage with and without the communication delay. Without the delay the over voltage lasts around 12 s and with the delay around 50 s. Here, another advantage of CHIL tests is seen. Due to the real-time execution it is also possible to study how timing issues (e.g., time delays) affects the performance of the controller.

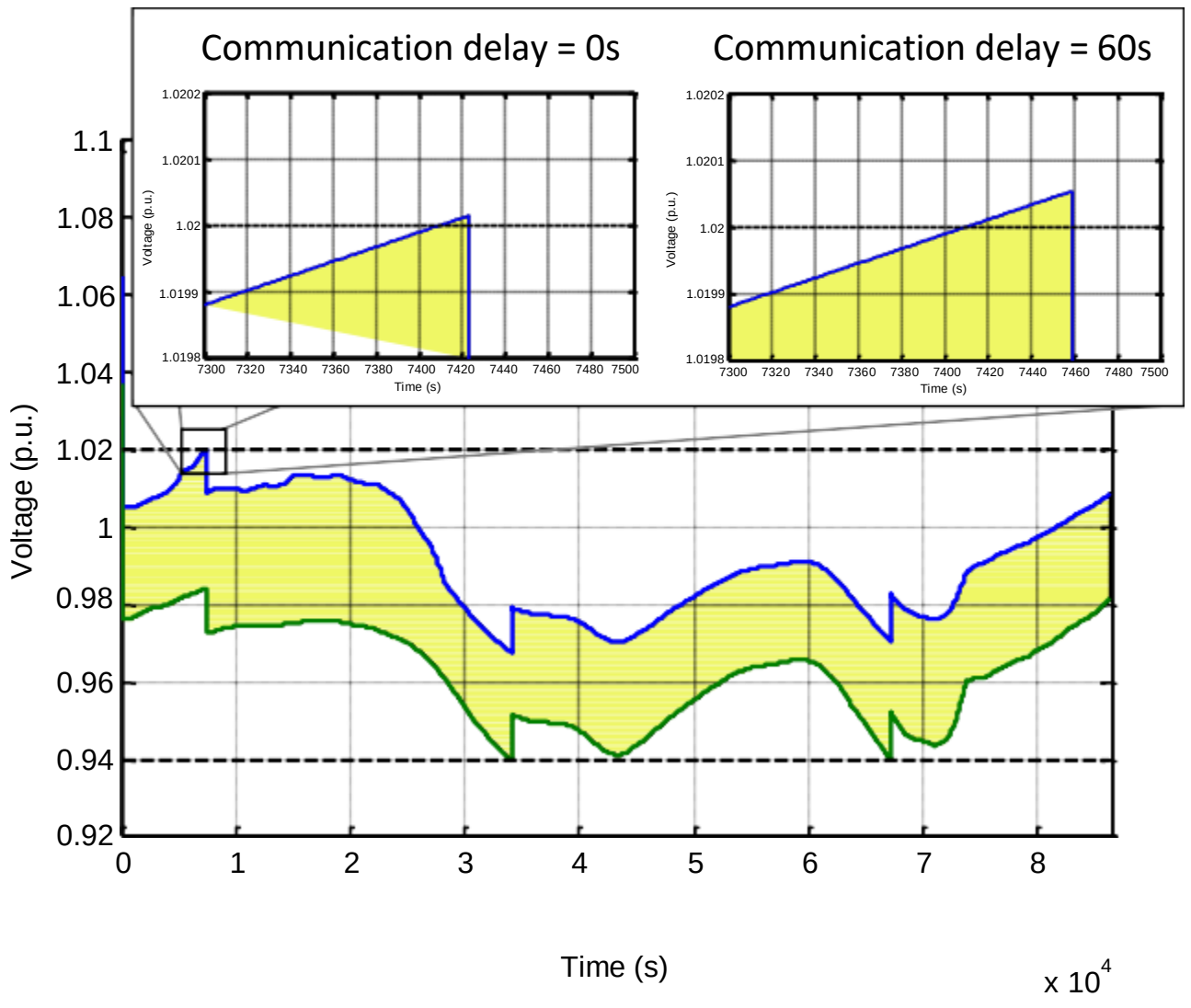

Fig. 7. Real-time simulation of the voltage band with the level control of the CVCU active.

## C. Validation of laboratory automation system

A further example shows the usage of CHIL simulation methodology for validating a laboratory automation approach [19]–[21]. The goal of this test is to validate the control application for the automatic tuning of an oscillatory circuit for PV-inverter anti-islanding detection capability tests in a CHIL environment. For this experiment the current from the grid to the PV-inverter has to be set to zero by automatically adjusting a local load (i.e., RLC circuit). This RLC circuit consists of a number of fixed loads which can be switched on or off (i.e., a coarse tuning) and a controllable load which is used for fine tuning.

Fig. 8 provides an overview of the realized CHIL setup. The laboratory environment is simulated using a DRTS. The automatic tuning algorithm for the oscillatory circuit is executed on an industry PC representing the controller. In addition, a SCADA system monitors measurement and control signals and provides an interface to the test engineer.

A sampling time of 1 ms was chosen for the execution of the CHIL test in order to have sufficient sample steps for the sine wave period of 20 ms in the execution of the model. This will ensure proper measurement in the modeled power meters.

The achieved test results of the CHIL experiment are provided in Fig. 9. The initial value of the PV-inverter model has been set to 5800 W. The coarse tuning algorithm gets triggered and reduces the active power into the grid to a value of 550 W. At this point the R-fine tuning starts increasing the load so that the active power into the grid approaches zero. After detecting the zero value for the active power into the grid the fine tuning stops and the test cycle is finished.

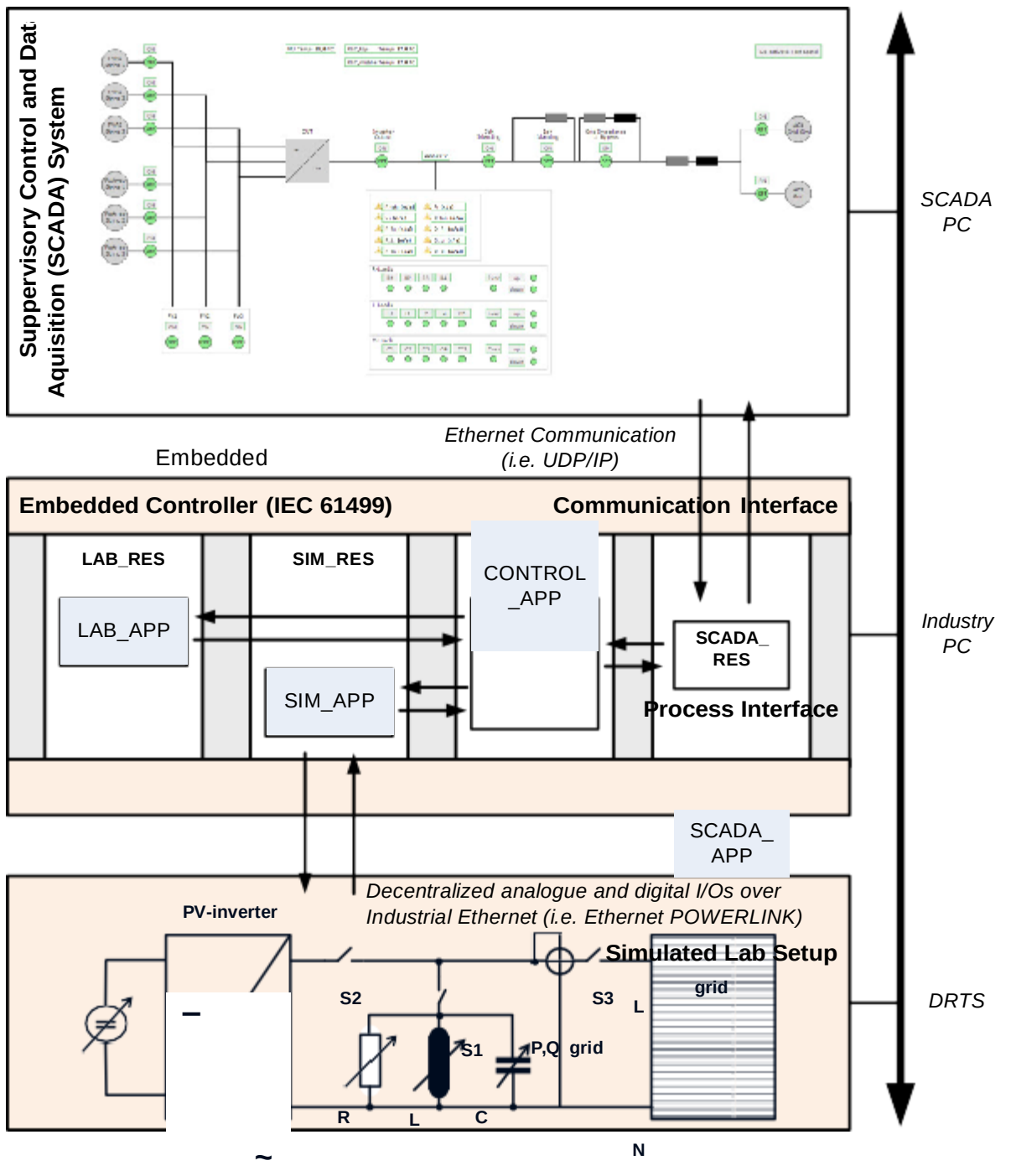

Fig. 8. CHIL setup for validating the lab automation (modified from [21]).

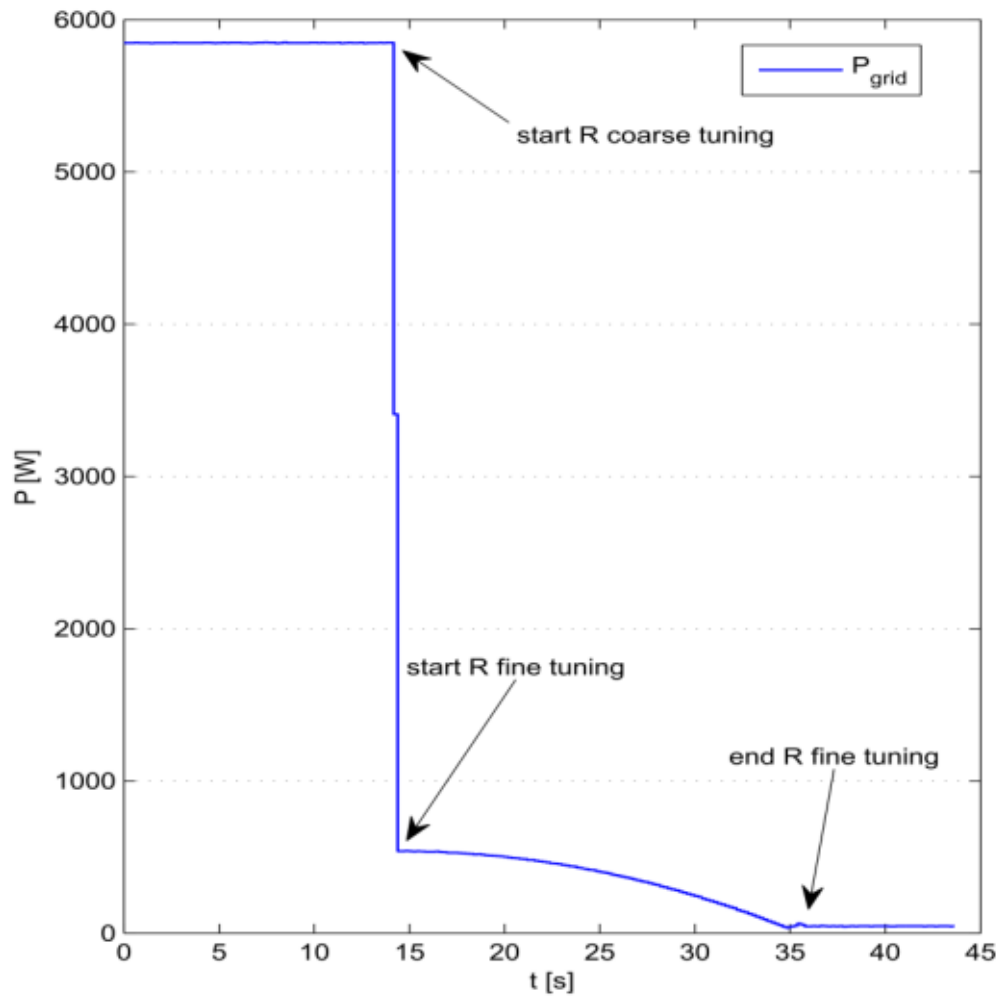

Fig. 9. CHIL simulation results of the lab automation application [21].

The test of the tuning of the local load was successful also in the real time simulation. The control application adjusted the active power into the grid to a value close to zero within the tolerance band of $\pm 3\%$.

Thus, the correct functionality of the lab controller application was tested in advance using the CHIL approach before it was implemented without any significant changes in the lab environment.

## IV. Future Standardization Needs

Some existing grid codes such as the guideline FWG BDEW TR4/8 [22], [23] additionally require validation by separate modeling of the full test setup in addition to the real laboratory testing (TR3). This modeling is currently done by creating software implementations of the grid, PV generation, HUT and interconnections. This also requires implementations of, e.g., load-flow and signal processing simulations. The latter implementations are a maintenance burden and more importantly could contain errors in the areas of modeling, simulation and code semantics.

The CHIL testing approaches mentioned earlier could both reduce implementation and maintenance costs for these mentioned supporting implementations. In CHIL testing, signal processing and load-flow simulations happen on real hardware so only the modeling of components connected to the HUT, e.g., grid and PV generation, is needed. Additionally, specific hardware platform requirements could make testing more standardized and reproducible compared to the current “all-software” approach.

The advantages of reduced development cost due to faster and simpler modeling as well as the possibility of reproducibility of modeling in validation requirements motivate the hypothesis that future standardization efforts will explicitly contain CHIL simulations as requirements. A high level automation of test sequences represents a crucial benefit for an improved testing process with the CHIL simulation methodology. In addition, different parameter settings can be run in parallel on a CHIL platform, hence the total testing time can be shortened due to individual acceleration features and parallel computing strategies.

A counter argument of using commercial load-flow software products for the validation model could be made. This would eliminate the development and maintenance burden mentioned above, but would also come with additional license costs, vendor lock-in to the respective commercial development environments and make modeling and simulation errors more non-transparent and harder to debug. Reproducibility in the context of commercial load-flow software products is made harder by version compatibility issues and integration costs. For these reasons it is assumed that the current practice of using commercial load-flow software products will change in favor of CHIL approaches.

## V. Conclusions

The work of this paper discussed the benefits of CHIL simulation and explains implementations applied for generator grid-compliance testing as well as voltage stability for networks. CHIL simulation systems enable automated testing procedures which increase the efficiency of the test duration. The flexibility of CHIL simulation setups is a major advantage compared to classical laboratory testing, because multiple parameter settings representing different hardware peripheries or network topologies can be executed by doing so. Another benefit is the generation of a fully automated test report which is of significant value as it is compliant with the reporting requirements of the selected standard.

A comparison of the obtained results from CHIL simulation and laboratory tests with PV inverters as physical hardware devices operated at rated power shows that the achievable accuracy within the CHIL platform is very high. In principle, an in depth-model of the power electronic circuitry and its periphery allows the CHIL simulation to produce identical results compared to a classical laboratory test setup. This fact is an encouraging result which can push advanced testing methods more into the focus of rapid prototyping and opens possibilities for future consideration and significance in major standardization frameworks.

The CHIL methodology enables the iterative development of smart grid related generation units such as grid-connected PV inverters. Design verification testing happens in a closed-loop HIL environment in which changes to controller logic and behavior can be interactively verified for a given setup. Automated report generation according to relevant standards completes the end-to-end testing from controller code through HIL simulation up to reports compliant to intended standards. The achieved degree of automation enables reproducible, scalable and transparent development for fast development cycles and time savings. Conducting pre-certification simulated testing upfront can cut costs on laboratory and staff later in the product development process.

## Acknowledgment

This work is partly supported by the European Communitys Horizon 2020 Program (H2020/2014-2020) under project "ERIGrid" (Grant Agreement No. 654113). The participation of AIT within ISGAN-SIRFN is funded in the frame of the IEA Research Cooperation program by the Austrian Ministry for Transport, Innovation and Technology (Contract No. FFG 839566).

This work is related to the international task force IEEE PES TF "Real-Time Simulation of Power and Energy Systems" and the international working group IEEE WG P2004 "Recommended Practice for Hardware-in-the-Loop (HIL) Simulation Based Testing of Electric Power Apparatus and Controls".